\documentclass[a4paper]{jpconf}

\usepackage[english]{babel} 

\usepackage[%
	final,%
]{graphicx}

\usepackage[
   centertags, % (default) center tags vertically
   sumlimits,  %(default) Place the subscripts and superscripts of summation
   intlimits,  % Like sumlimits, but for integral symbols.
   namelimits, % (default) Like sumlimits, but for certain 'operator names' such as
]{amsmath} %
\usepackage{amssymb}

\usepackage[%
	pdftitle={Production and elliptic flow of heavy quarks at RHIC and LHC},%
	pdfauthor={Jan Uphoff, Oliver Fochler, Zhe Xu, and Carsten Greiner},%
	pdfsubject={Production and elliptic flow of heavy quarks at RHIC and LHC},
	pdfstartview=FitH,
	pdfpagemode=UseNone,
	bookmarksopen=true
	]{hyperref}

\usepackage[all]{hypcap} % Links to figures point actually on figure itself, not on caption

\usepackage{fancyhdr}

\usepackage{dcolumn} % tabellenasurichtung am komma

\usepackage{multirow} % tabelleneinträge über mehrere Zeilen

\usepackage{microtype} % bessere trennung

\makeatletter
\def\tagform@#1{\maketag@@@{\ignorespaces#1\unskip\@@italiccorr}}
\let\orgtheequation\theequation
\def\theequation{(\orgtheequation)}
\makeatother

\newcommand{\beq}{\begin{equation}}
\newcommand{\eeq}{\end{equation}}
\newcommand{\umbruch}{\nonumber \\}
\newcommand{\gnuplotwidth}{0.6\textwidth}

\begin{document}
\title{Production and elliptic flow of heavy quarks at RHIC and LHC within a partonic transport model}

\author{Jan Uphoff,$^1$ Oliver Fochler,$^1$ Zhe Xu,$^{2,1}$ and Carsten Greiner$^1$}

\address{$^1$ Institut f\"ur Theoretische Physik, Johann Wolfgang 
Goethe-Universit\"at Frankfurt, Max-von-Laue-Str. 1, 
D-60438 Frankfurt am Main, Germany}
\address{$^2$ Frankfurt Institute for Advanced Studies, Ruth-Moufang-Str. 1, D-60438 Frankfurt am Main, Germany}

\ead{uphoff@th.physik.uni-frankfurt.de}

\begin{abstract}
Production and elliptic flow of heavy quarks are investigated in nucleus-nucleus collisions at RHIC and LHC within the \emph{Boltzmann Approach of MultiParton Scatterings} (BAMPS). The initial heavy quark yield is estimated with the leading order mini-jet model and compared to PYTHIA for several parton distribution functions. Secondary production of heavy quarks in the quark-gluon plasma is examined within the full 3+1 dimensional BAMPS simulation of heavy ion collisions. At RHIC this yield is negligible, but for LHC charm production in the QGP plays a significant role. In addition, we study the elliptic flow of charm quarks at RHIC with the result that leading order processes are not sufficient to describe the experimental data.

\end{abstract}

\section{Introduction}
Charm and bottom quarks  are an interesting probe for the hot and dense medium, which is produced in heavy ion collisions at RHIC \cite{Adams:2005dq,Adcox:2004mh,Arsene:2004fa,Back:2004je}. Due to their large mass heavy quarks are produced at a very early stage of the collision, revealing 	unique insight into the initial properties of the quark-gluon plasma (QGP). This large mass also determines the relevant scale of their production processes, which should, therefore, be describable in the framework of perturbative QCD (pQCD).

The production of heavy quarks can be divided in three stages: They are created (i) in hard scatterings during initial nucleon-nucleon collisions, (ii) in the QGP and (iii) during the hadronic phase. We will neglect the latter one since the energy density in the hadronic gas is low and heavy flavor production becomes very unlikely.  At RHIC and LHC primary heavy quark production in hard scatterings plays the most important role. At RHIC secondary production in the QGP is nearly negligible, but at LHC it contributes significantly to the total heavy flavor yield.

The experimentally observed elliptic flow \cite{Adare:2006nq} and energy loss \cite{Abelev:2006db,Adare:2006nq} of heavy quarks indicate that they interact strongly with the medium and thermalize relatively fast, although the ``dead cone effect'' \cite{Dokshitzer:2001zm,Zhang:2003wk} predicts a delay in their thermalization compared to light quarks. The investigation of this puzzle is still in progress \cite{Armesto:2005mz,vanHees:2005wb,Moore:2004tg,Wicks:2005gt,Peigne:2008nd,Gossiaux:2008jv}.

The present article is organized as follows: In the next section we describe our model, the parton cascade BAMPS. Thereafter, we discuss primary heavy quark production in hard parton scatterings during initial nucleon-nucleon collisions within the leading order mini-jet model and PYTHIA. In \autoref{sec:prod_qgp}
%Sec. \ref{sec:prod_qgp} !!
secondary heavy quark production in the QGP at RHIC and LHC is investigated with BAMPS for different initial conditions. Finally, we address elliptic flow of charm quarks at RHIC in \autoref{sec:elliptic_flow}, followed by brief conclusions.

\section{Parton cascade BAMPS}
\label{sec:bamps}

The full $3+1$ space-time evolution of the QGP is studied within the partonic transport model \emph{Boltzmann Approach of MultiParton Scatterings} (BAMPS) \cite{Xu:2004mz,Xu:2007aa}, which solves the Boltzmann equation,
\begin{equation}
\label{boltzmann}
\left ( \frac{\partial}{\partial t} + \frac{{\mathbf p}_i}{E_i}
\frac{\partial}{\partial {\mathbf r}} \right )\, 
f_i({\mathbf r}, {\mathbf p}_i, t) = {\cal C}_i^{2\rightarrow 2} + {\cal C}_i^{2\leftrightarrow 3}+ \ldots  \ ,
\end{equation}
dynamically for on-shell partons with a stochastic transport algorithm and pQCD interactions. ${\cal C}_i$ denotes the collision integrals of the $2\rightarrow 2$ and also $2\leftrightarrow 3$ interactions  and $f_i({\mathbf r}, {\mathbf p}_i, t)$ the one particle distribution function of species $i=g,\, Q,\, \bar{Q}$ ($Q=c,b$), since light quarks are not included yet. The following processes are implemented in BAMPS:
\begin{align}
\label{bamps_processes}
	g+g &\rightarrow g+g \nonumber\\
	g+g &\rightarrow g+g+g \nonumber\\
	g+g+g &\rightarrow g+g	 \nonumber\\
	g+g &\rightarrow Q +\bar{Q} \umbruch
	Q+ \bar{Q} &\rightarrow g+g \umbruch
	g+Q &\rightarrow g+Q \nonumber\\
	g+\bar{Q} &\rightarrow g+\bar{Q}
\end{align}

For further details on the purely gluonic processes  we refer to \cite{Xu:2004mz,Xu:2007aa}  and to \cite{Uphoff:2010sh} for the processes, in which heavy quarks are involved.

\section{Primary heavy quark  production in initial nucleon-nucleon scatterings}
\label{sec:ini_charm}

For the estimation of the primary heavy quark yield during initial nucleon-nucleon scatterings we use the following two approaches.

\subsection{Leading order heavy quark  production within the mini-jet model}
\label{sec:charm_lo}

In leading order (LO) heavy quarks are produced in the two processes
\begin{align}
\label{charm_prozesse}
	g+g &\rightarrow Q +\bar{Q} \nonumber \\
	q+ \bar{q} &\rightarrow Q +\bar{Q}  \ .
\end{align}

The differential cross section for heavy quark production in the collision of hadron $A$ and $B$ is essentially given by the product of the parton distribution functions (PDF) $f_i$ and the differential partonic cross sections for both processes, \cite{Gavai:1994gb}
\begin{align}
\label{pp_cs_dist}
\frac{\mathrm{d} \sigma_{Q\bar{Q}}^{AB}}{\mathrm{d}p_T^2 \mathrm{d}y_Q \mathrm{d}y_{\bar{Q}}} &= x_1 x_2
\Biggl[ f^A_g(x_1) \, f^B_g (x_2) \, \frac{\mathrm{d}\hat{\sigma}_{gg \rightarrow Q \bar{Q}}}{\mathrm{d}\hat{t}}   \nonumber \\
&\quad\quad\quad\quad   + 
	\sum_q \left[ f^A_q(x_1) \, f^B_{\bar{q}} (x_2) + f^A_{\bar{q}}(x_1) \, f^B_q (x_2) \right] \frac{\mathrm{d}\hat{\sigma}_{q\bar{q} \rightarrow Q \bar{Q}}}{\mathrm{d}\hat{t}}  \Biggr] \ ,
\end{align}
$y$ being the rapidity and $p_T$ the transversal momentum of the heavy quark and anti-quark in the center of mass frame. 
The PDFs are dependent on the factorization scale and the partonic cross sections on the renormalization scale. Therefore, heavy quark production in general is sensitive to both scales. Other uncertainties come from the heavy quark mass and the choice of PDFs. Depending on these parameters the total LO heavy quark production cross section in nucleon-nucleono collisions changes by a factor of about 2 at RHIC and even of about 4 at LHC energy \cite{Uphoff:2010sh}.

\subsection{Heavy quark production with PYTHIA}
\label{sec:ini_charm_pythia}

In addition, we use PYTHIA \cite{Sjostrand:2006za}, the LO event generator for nucleon-nucleon collisions, to estimate the number of initially produced heavy quarks in heavy ion collisions. For that we scale heavy quarks from nucleon-nucleon collisions with roughly the number of binary collisions\footnote{Actually, this value is lowered to $1000$ at RHIC \cite{Adams:2004cb} and $1500$ at LHC \cite{Emel'yanov:1999bn,Accardi:2004be} in order to take shadowing into account.}. Details on how this is done can be found in \cite{Uphoff:2010sh}.
Tables \ref{tab:pythia_charm_yield} and \ref{tab:pythia_charm_yield_lhc} list the number of initially produced charm quarks in central Au+Au collisions at RHIC and charm and bottom quarks in central Pb+Pb collisions at LHC, respectively, according to PYTHIA for various parton distribution functions. As a note, there are also large errors due to uncertainties in mass, shadowing, factorization and renormalization scale like in the LO pQCD calculation, although these are not reflected in the tables.
\begin{table}
	\centering
		\begin{tabular}{l|c|D{!}{.}{5.4}}
			PDF & Reference &  
			\multicolumn{1}{c}{Charm pairs}			\\ \hline
			\hline
CTEQ5l (LO) & \cite{Lai:1999wy} & 8!9 \\ \hline
CTEQ6l (LO) & \cite{Pumplin:2002vw_CTEQ6} & 9!2 \\ \hline
CTEQ6m ($\overline{MS}$) & \cite{Pumplin:2002vw_CTEQ6} & 13!6 \\ \hline
%MRST2001LO & \cite{Martin:2002dr} & 9!6 \\ \hline
MRST2007LOmod & \cite{Sherstnev:2007nd} & 9!2\\ \hline
HERAPDF01 & 
\cite{Nagano:2008ip} 
 & 12!3 \\ \hline
%GJR08 (FF LO) & \cite{Gluck:2007ck,Gluck:2008gs} & 3!0 \\ \hline
GRV98 (LO) & \cite{Gluck:1998xa} & 3!0
		\end{tabular}
	\caption{Number of charm pairs produced in primary hard scatterings in central Au+Au collisions at RHIC for some parton distribution functions by sampling nucleon-nucleon collisions with PYTHIA and scaling to Au+Au collisions.}
	\label{tab:pythia_charm_yield}
\end{table}
\begin{table}
	\centering
		\begin{tabular}{l|c|c|c} 
		&& \multicolumn{2}{c}{Heavy quark pairs}\\ 
			PDF & Reference & 
			Charm  & Bottom			\\ \hline
			\hline
CTEQ6l (LO) & \cite{Pumplin:2002vw_CTEQ6} & 62 & 7.2 \\ \hline
CTEQ6m ($\overline{MS}$) & \cite{Pumplin:2002vw_CTEQ6} & 66 & 6.9 \\ \hline
MRST2007LOmod & \cite{Sherstnev:2007nd} & 67 & 8.9
		\end{tabular}
	\caption{As \autoref{tab:pythia_charm_yield} but for central Pb+Pb collisions at LHC.}
	\label{tab:pythia_charm_yield_lhc}
\end{table}

\begin{figure}
	\centering
\includegraphics[width=\gnuplotwidth]{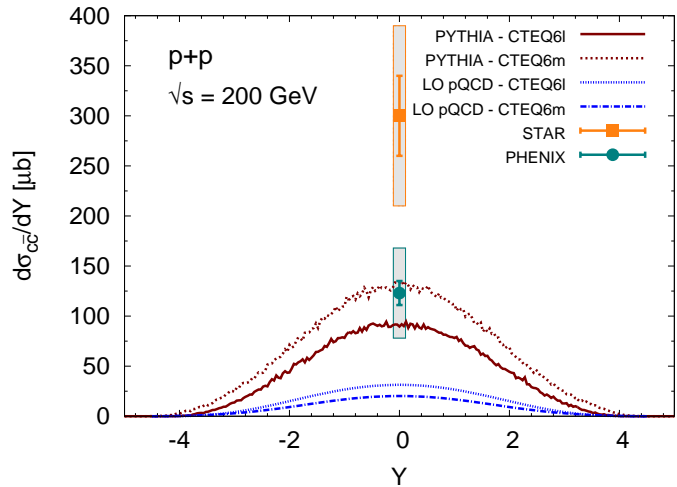}%eps
	\caption{Charm production cross section $\mathrm{d}\sigma_{c \bar{c}}^{NN}/\mathrm{d}y$ as a function of rapidity $y$ in a nucleon-nucleon collision at RHIC energy simulated with PYTHIA and pQCD, respectively, for the PDFs CTEQ6l and CTEQ6m together with experimental data \cite{Adams:2004fc_STAR_dcsdY_cstot,Adare:2006hc_PHENIX_dsigmadY}. The pQCD calculation is done in LO with $\mu_F = \mu_R=\sqrt{p_T^2+M_c^2}$, $N_f=3$, $M_c= 1.5\, {\rm GeV}$, $\lambda_{\rm QCD}=346 \, \rm{MeV}$ \cite{Bethke:2006ac} and $K=1$.}
	\label{fig:ini_charm_dn_dy_exp}
\end{figure}

\autoref{fig:ini_charm_dn_dy_exp} compares the rapidity distribution of the charm production cross section in one nucleon-nucleon collision simulated with PYTHIA together with the LO pQCD calculations and the experimental data points \cite{Adams:2004fc_STAR_dcsdY_cstot,Adare:2006hc_PHENIX_dsigmadY}.
The LO calculations are far below both experimental values, which indicates that next-to-leading order processes \cite{Cacciari:2005rk} have to be considered. In contrast, PYTHIA lies much closer at least to the PHENIX data point. Therefore, we decided to use PYTHIA for sampling our initial heavy quark distributions. For the PDFs we chose CTEQ6l, because it is designed for LO event generators such as PYTHIA \cite{Pumplin:2002vw_CTEQ6}.

\section{Secondary heavy quark production in the QGP}
\label{sec:prod_qgp}

\subsection{Charm production at RHIC}
\label{sec:prod_qgp_rhic}

We study charm production during the QGP phase in central Au+Au collisions at RHIC within the framework of BAMPS. The initial charm distribution is obtained from PYTHIA. Initial gluons are sampled according to three different models: a color glass condensate (CGC) inspired model \cite{Drescher:2006pi,Drescher:2006ca}, the mini-jet model \cite{Kajantie:1987pd,Eskola:1988yh}, and PYTHIA. The detailed prescription how we scale partons from PYTHIA to heavy ion collisions can be found in \cite{Uphoff:2010sh}. For the partonic cross sections in BAMPS of the processes from \ref{bamps_processes} we employ a constant coupling of $\alpha_s = 0.3$.

\begin{figure}
	\centering
\includegraphics[width=\gnuplotwidth]{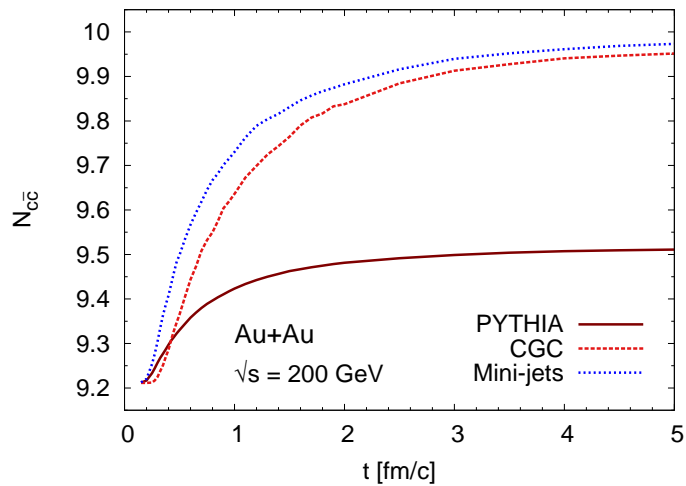}%eps
	\caption{Number of charm quark pairs produced in a central Au+Au collision at RHIC according to BAMPS. The initial parton distribution for gluons is obtained with PYTHIA, CGC, and the mini-jet model. For better comparison the initial charm distribution from PYTHIA is used for all models. Please note the small range of the $y$ axis.}
	\label{fig:charm_yield_rhic_central_initial}
\end{figure}

\autoref{fig:charm_yield_rhic_central_initial} shows the charm yield in the QGP as a function of time.
For all three initial distribution models the charm production during the QGP is very small compared to the initial yield. With PYTHIA initial conditions only 3\,\% of the total final charm quarks are produced in the QGP. For CGC or mini-jet initial distributions for the gluons this fraction is about 6\,\%. The higher yield in the latter models is due to the higher initial energy density \cite{Uphoff:2010sh}. The main increase in the charm number takes place at an early stage of the evolution, where the QGP is not fully equilibrated yet. 

\begin{figure}
	\centering
\includegraphics[width=\gnuplotwidth]{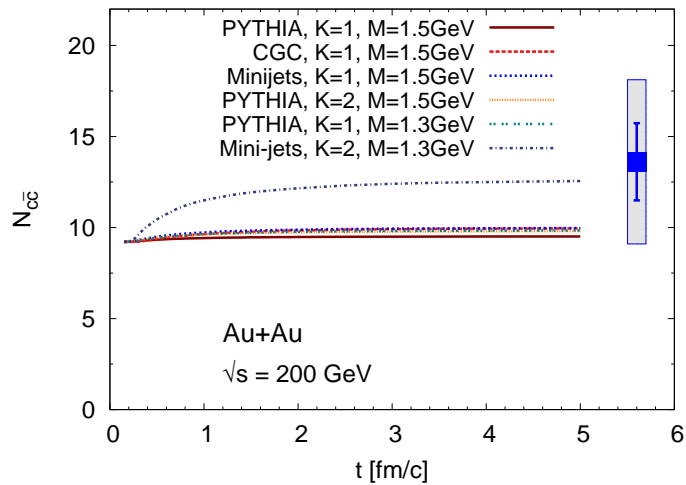}%eps
	\caption{As \autoref{fig:charm_yield_rhic_central_initial}, varying initial gluon conditions, charm mass $M$ and $K$ factor for  $gg \rightarrow c \bar{c}$. In addition, the experimental value for the number of final charm pairs is plotted \cite{Adler:2004ta}.}
	\label{fig:charm_yield_rhic_central_all}
\end{figure}

Changing the charm mass to a smaller value of $M_c = 1.3\, {\rm GeV}$ or employing a $K$ factor of 2 to the charm production cross section increase these results by a factor of 2, respectively. As a result, using mini-jet initial conditions, $M_c = 1.3\, {\rm GeV}$ and also  $K=2$ gives the maximum value in our model of 3.4 charm pairs produced in the QGP (cf. \autoref{fig:charm_yield_rhic_central_all}), which is still just 27\,\% of the total charm yield.

Our results indicate that the charm production in the QGP at RHIC is nearly negligible. That is also underlined by experiment, where a charm scaling with the number of binary collisions was found \cite{Adler:2004ta,Adams:2004fc_STAR_dcsdY_cstot,:2008hja}. Charm quarks are, therefore, produced in initial hard scatterings during nucleon-nucleon collisions.

At first sight this small charm production is a bit surprising since the charm quark fugacity is always smaller than 1 in the QGP \cite{Uphoff:2010sh}. However, the reason for that lies in the huge timescale of chemical equilibration of charm quarks in the medium produced at RHIC, which is about a factor of 100 larger than the lifetime of the QGP \cite{Uphoff:2010sh}.

\subsection{Charm and Bottom production at LHC}
\label{sec:prod_qgp_lhc}
\begin{figure}
	\centering
\includegraphics[width=\gnuplotwidth]{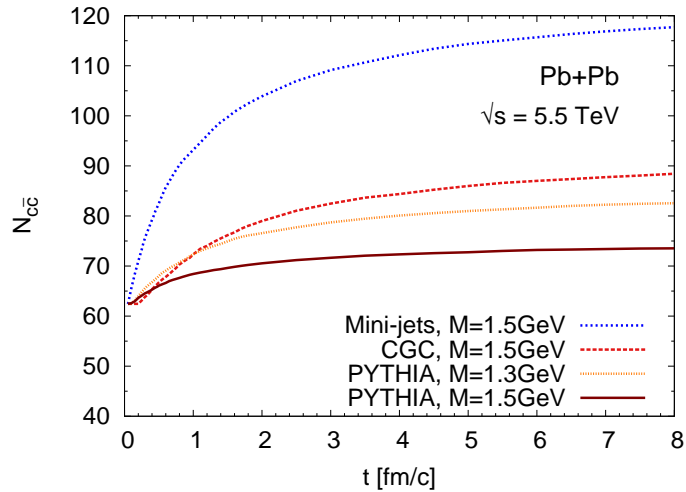}%eps
	\caption{Number of charm pairs in a central Pb+Pb collision at LHC simulated with BAMPS for PYTHIA, CGC, and mini-jet initial conditions for gluons. Charm quarks are sampled with PYTHIA for all three models.}
	\label{fig:charm_yield_lhc_central}
\end{figure}

At LHC the picture changes compared to RHIC: Due to the larger energy density, charm production in the QGP cannot be neglected anymore. \autoref{fig:charm_yield_lhc_central} depicts our predictions for the charm number evolution at LHC. Again, we use PYTHIA for the initial charm yield and PYTHIA, CGC, and mini-jet  initial conditions for the gluons.
With PYTHIA initial conditions 15\,\% of the total charm quarks are produced in the QGP. With gluons from CGC this fraction even rises and for mini-jet  initial gluon conditions the charm yield during the QGP phase is nearly equal to the initial charm number. Lowering the charm mass or introducing a $K$ factor increase these values again.

Just like at RHIC, the charm quark fugacity at LHC is far below 1 \cite{Uphoff:2010sh}. However, because of the much smaller equilibration time of charm quarks in such a hot medium as at LHC, which is of the same order of magnitude as the QGP lifetime \cite{Uphoff:2010sh}, the charm production is much stronger.

As shown in \cite{Uphoff:2010sh} the bottom production during the QGP phase at LHC and, of course, also at RHIC is completely negligible due to the larger mass. Therefore, bottom quarks are just produced in initial hard scatterings, which renders them as a promising probe of the early stage of the collision due to the precisely determined production time.

\section{Elliptic flow of charm quarks at RHIC}
\label{sec:elliptic_flow}

\begin{figure}
	\centering
\includegraphics[width=\gnuplotwidth]{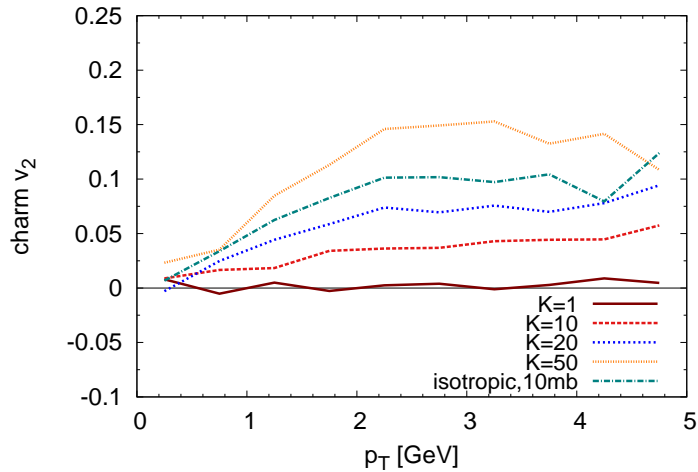}%eps
	\caption{Elliptic flow of charm quarks with pseudo-rapidity $|\eta|<0.35$ at time $t=5 \, {\rm fm/c}$ for RHIC collisions with impact parameter of $b=8.2 \, {\rm fm}$. The cross section of $gc \rightarrow gc$ is multiplied with different $K$ factors.}
	\label{fig:charm_v2_rhic_k}
\end{figure}

The elliptic flow of charm quarks is defined by the average 
\begin{align}
\label{elliptic_flow}
	v_2=\left\langle  \frac{p_x^2 -p_y^2}{p_T^2}\right\rangle 
\end{align} 
over  all charm quarks at mid-rapidity and is an indicator how strong charm quarks are coupled to the bulk medium. In the present article we study only charm scattering with gluons in leading order, $g+c \rightarrow g+c $. Without considering higher order corrections or employing a $K$ factor for the cross section the interaction between charm quarks and the gluonic medium is rather small. Therefore, although a strong gluonic $v_2$ is observed in BAMPS \cite{Xu:2007jv,Xu:2008av,Xu:2010cq}, which is in agreement with experimental data, the $v_2$ of charm quarks with $K=1$ vanishes nearly, as shown in \autoref{fig:charm_v2_rhic_k}.
For a substantial charm flow one must multiply the cross section of $gc \rightarrow gc $ by at least a factor of 10. In order to come in the region of the experimentally observed flow of heavy flavor \cite{Adare:2006nq,Dion:2009} even an unreasonable $K=50$ is necessary, which shows that the charm scattering cannot be described just by the LO process. Of course, one has to be careful with comparing the flow from charm quarks to the experimentally observed flow of heavy flavor electrons, which stem from the decay of $D$ or $B$ mesons. However, in \cite{Zhang:2005ni} it is found that the effect of hadronization and decay on the flow is negligible. In contrast, in the coalescence model the flow of $D$ mesons is considerably larger than that of charm quarks \cite{Molnar:2004ph}.
We also made some calculations with bottom quarks and saw that our here presented results are only slightly modified and that only for high $p_T$, if one considers both charm and bottom.

In order to improve our calculations we will investigate the effect of a running coupling and modify the evaluation of the screening mass in the $gc \rightarrow gc $ cross section \cite{Peshier:2008bg,Peigne:2008nd,Gossiaux:2008jv,Gossiaux:2009mk}. In addition, we want to study the impact of taking higher order corrections like $g+c \rightarrow g+c+g$ into account, for which BAMPS is an ideal framework since $2 \leftrightarrow 3$ interactions are already included for gluons \cite{Xu:2004mz}. With these improvements we will also extend recent studies on the nuclear modification factor \cite{Fochler:2008ts,Fochler:2010wn} to the heavy flavor sector.

\section{Conclusions}

We have studied heavy quark production in heavy ion collisions at RHIC and LHC. The initial yield was estimated with LO pQCD processes, which did not agree with the data, and PYTHIA. The latter was then used to sample the initial heavy quark distribution for the QGP phase, which was simulated with BAMPS. In that framework we investigated the dependence of heavy quark production on different initial gluon distributions, namely from PYTHIA, CGC and the mini-jet model.

At RHIC the production of charm and, of course, bottom quarks in the QGP is highly suppressed due to the huge chemical equilibration time scale in the medium created at RHIC. For LHC we predict a significant charm production of 10 to 60 charm pairs in the QGP phase, depending on the initial conditions of the medium. This is due to the high initial energy density and a relatively small chemical equilibration time scale, which is of the same order as the QGP lifetime. In contrast, bottom production in the QGP at LHC is negligible due to the high bottom mass.

The experimentally observed elliptic flow of heavy quarks cannot be described with LO pQCD cross sections. Therefore, we want to take higher orders into account and study the effect of a running coupling and improved Debye screening.

\section*{Acknowledgements}
The BAMPS simulations were performed at the Center for Scientific Computing of the Goethe University. This work was supported by the Helmholtz International Center for FAIR within the framework of the LOEWE program (Landes-Offensive zur Entwicklung Wissenschaftlich-\"okonomischer Exzellenz) launched by the State of Hesse.
We would like to thank the organizers of the 26th Winter Workshop on Nuclear Dynamics for this pleasant and stimulating conference.

\section*{References}
\bibliography{hq}

\end{document}